  \documentclass[final,preprintnumbers]{revtex4}
 \usepackage{amsmath} 
\usepackage{amsfonts} 
\usepackage{amssymb}
\usepackage{graphicx} 

\begin{document}

\title{Fundamental Constants\footnote{Contribution to ``Visions of Discovery'' in honor of Charles Townes' 90th Birthday}}
\author{Frank Wilczek}
\vspace*{.2in}
\affiliation{Center for Theoretical Physics \\
Department of Physics, Massachusetts Institute of Technology\\
Cambridge Massachusetts 02139 USA}
\preprint{MIT/CTP-3847}
\vspace*{.3in}

\begin{abstract}
The notion of ``fundamental constant'' is heavily theory-laden.   A natural, fairly precise formulation is possible in the context of the standard model (here defined to include gravity).   Some fundamental constants have profound geometric meaning.  The ordinary gravitational constant parameterizes the stiffness, or resistance to curvature, of space-time.  The cosmological term parameterizes space-time's resistance to expansion -- which may be, and apparently is at present, a {\it negative\/} resistance, i.e. a tendency toward expansion.   The three gauge couplings of the strong, electromagnetic, and weak interactions parameterize resistance to curvature in internal spaces.    The remaining fundamental couplings, of which there are a few dozen, supply an ungainly accommodation of inertia.   The multiplicity and variety of fundamental constants are  esthetic and conceptual shortcomings in our present understanding of foundational physics.   I discuss some ideas for improving
the situation.   I then briefly discuss additional constants, primarily cosmological, that enter into our best established present-day world model.  Those constants presently appear as macroscopic state parameters, i.e. as empirical ``material constants'' of the Universe.  I mention a few ideas for how they might become fundamental constants in a future theory.   In the course of this essay I've advertised several of my favorite speculations, including a few that might be tested soon.   
\end{abstract}

\maketitle

\newpage

Our quest for {\it precise\/} correspondence between mathematical equations and real-world phenomena separates modern physics and its allied fields from any other intellectual enterprise, past or present.    Through his path-breaking work in molecular spectroscopy, Charles Townes greatly extended the scope and refined the precision of that correspondence.   Rigorous mathematical formulations of reality support long lines of logical deductions, and thereby open up opportunities for {\it designing surprises}.   Townes'  pioneering design of masers and lasers are outstanding examples of that special variety of creativity.    

\bigskip

A correspondence is more powerful, the more each side of it can function independently from the other.  On the mathematical side of the equations-reality correspondence, the ideal result would be a model defined purely conceptually, i.e. without any explicit reference to phenomena.   The phenomena would then be derived in their entirety.   

Physics has come far toward achieving that goal, but has not reached it.   One objective measure of the remaining distance is that our equations contain purely numerical quantities -- parameters -- whose values cannot be derived from those equations.  In the correspondence between equations and reality, the values of those parameters are not determined conceptually, but must be provided empirically, by measurements.    In this way, we introduce ``fundamental constants''  into our description of Nature.  

\section{Preliminary: Fundamental Constants and Systems of Units}

What is a fundamental constant, exactly?  I don't think  there's a precise, universally agreed answer to that question.  Genuine subtleties surround the concept, that have led to inconclusive debates \cite {Sommerfeld}, \cite{Duff}.   I'll start this essay by crafting and defending a reasonably precise definition. In the process I'll discuss some of the subtleties involved, and motivate my choices for resolving some ambiguities.

\subsection{Units and Assumptions}
 
Numerical values for measured quantities such as lengths or times are often obtained only after they are compared to reference values -- {\it e.g}. six meters, 1.34 seconds.   In other words, we require a system of units.  Establishing units is an important part of defining the context in which we discuss fundamental constants.  So preliminary to a critical definition and enumeration of fundamental constants, it will be useful to address issues that surround the choice of a system of physical units.   

In Volume 3 of Arnold Sommerfeld's famous {\it Lectures on Theoretical Physics}, devoted to electrodynamics, articles 7 and 8 (out of a total of 38) are a lengthy discussion of the choice of a system of units.  Article 8 is entitled ``Four, Five, or Three Fundamental Units?''   A central issue, for Sommerfeld, is whether to include separate units for electric and magnetic charge, in addition to the standard three mechanical units for mass, length and time ($[M,L,T]$).   

If one chooses to introduce a separate unit $[Q]$ for charge, then one must introduce, in Coulomb's force law $F \propto \frac{q_1q_2}{r^2}$, a conversion factor mediating between the mechanical units $[MLT^{-2}]$ appearing on the left hand side and the different units $[Q^2L^{-2}]$ that appear on the right hand side.   That is accomplished, notoriously, by introducing a conversion factor $\epsilon_0$, the electric permeability of vacuum, so that $F = \frac{1}{4\pi \epsilon_0} \frac{q_1q_2}{r^2}$.    

Alternatively, one can use Coulomb's force law to define the unit of charge in terms of mechanical units.  The force between two unit charges, at a distance of one centimeter, is then one dyne, by definition.  

The first procedure, which regards quantity of charge as a separate concept, independent of its mechanical effects, appears natural if the concept {\it quantity of charge\/} has some other independent physical meaning.   In atomic physics, of course, we learn that it does.  There is an operationally defined, reproducible unit of electric charge, the charge of an electron.   We can express other charges as numerical multiples of that unit.   Then $\epsilon_0$ becomes a measurable ``fundamental constant'', parameterizing the Coulomb force between electrons.   In this system, the numerical value of the electron charge is unity, by definition: $e = 1\  [Q]$.   

Alternatively, in the second procedure, the same measurements would give us the (non-trivial) numerical value of the electron charge, expressed in purely mechanical units.    

Finally, we could follow Yogi Berra's advice: ``When you come to a fork in the road, take it.''  Combining the two procedures, we could express the electron charge in mechanical units, {\it and\/} set it equal to unity.   In this way, we set a certain combination of $[M, L, T]$, namely $[Q^2] = [ML^3T^{-2}]$, to unity.  Thus we reach a system that contains just two independent mechanical units.   

Of course there is no real mystery or objective dispute about the elementary physics under discussion here.  The different alternatives simply represent different ways of expressing the same facts.   Thus we see clearly that identifying fundamental constants, or even counting their number, involves an element of convention.   It depends, first of all, on how many units we choose to keep.   If we keep additional units (such as $[Q]$) then we will need additional fundamental constants (such as $\epsilon_0$) to mediate equations in which they appear.  More profoundly, it depends upon where we choose to draw the dividing line between facts so well established that we are comfortable to regard them, at least provisionally, as {\it    
a priori\/} features of our theoretical world-model, and issues we choose to keep open.   If we take the existence of electrons all having rigorously the same charge at all times and places as an established fact, then we can use the universal value of electron's charge as the unit of charge; if we do not, we must get the unit from elsewhere.   If we take the validity of Coulomb's law as an established fact, then we can use it to express charge in mechanical units; if not, we must keep an independent unit of charge.  (Assuming, of course, that we do not substitute some other electromechanical law for Coulomb's).    And if we assume both things, then we can both define charge in mechanical units {\it and\/} reduce the number of mechanical units.

In general, the more facts we allow ourselves to assume {\it a priori}, the fewer units, and the fewer fundamental constants, we need to introduce. Here is an example, to illustrate the point further.  Suppose that we chose not to assume that the equations of physics are rotationally invariant, but allowed for the possibility that they contained a preferred direction.  Then we could formulate two independent versions of Coulomb's law: one that applies when the line between the charges lies in the preferred direction, and another that applies when that line is transverse to it.  Each law would support its own $\epsilon$: $\epsilon_{\rm long.}$ and $\epsilon_{\rm trans.}$.   Measurements would, of course, establish the near equality of those two fundamental constants.  But as a matter of principle we can only remove the ``fundamental constant'' $\epsilon_{\rm long.}/\epsilon_{\rm trans.}$ by adopting a theoretical assumption.    In general, by being bold we'll be economical, and appropriately ambitious -- but we might be wrong.

\subsection{Defining Units and Fundamental Constants}

At present, and for the past 35 years or so, the irreducible laws of physics -- that is, the laws which we don't know how to derive from other ones -- can be summarized in the so-called standard model.   So the standard model appears, for the present, to be the most appropriate {\it a priori\/} context in which to frame the definition of fundamental constants.

Here and below I'll be using a slightly non-standard definition of the standard model, in two respects.  First, I include gravity, by means of Einstein's general relativity, implemented with a minimal coupling procedure.  Second, I include neutrino masses and mixings.   I'll come back to defend those inclusions below. 

The standard model is specified, in practice, by its Lagrangian.  Given the Lagrangian, we can derive the equations of our current best world-model,  and their physical interpretation, following the methods of relativistic quantum field theory.    (More precisely, the equations tell us what sorts of matter might exist, and how they will behave; they do not tell us what objects actually exist.  Or, in jargon: They tell how any state evolves in time, but not which particular state describes the world.)  In this framework, there is a clear and natural definition of what we mean by a fundamental constant.  A fundamental constant is a parameter whose value we must supply in order to specify the Lagrangian of the standard model.  

Since the principles of special relativity and quantum mechanics are deeply woven into the fabric of the standard model, it seems appropriate to define $c$ and $\hbar$ as the units of velocity and action, respectively.  In these units, of course, $\hbar = c = 1$, so those quantities do not appear explicitly as parameters.

Indeed, $c$ is strictly analogous to the parameter $\epsilon_{\rm long.}/\epsilon_{\rm trans.}$ we discussed above!  It is a parameter to accommodate possible differences in values between quantities whose relative values are fixed by symmetry.  While we can and should continue to test both Lorentz invariance and rotational invariance, it seems extravagant to carry around the excess baggage of parameters whose value is fixed by a symmetry until a violation of the symmetry is discovered, or at least plausibly suggested.   

$\hbar$ also appears in a symmetry algebra, although a much more abstract one, the algebra of canonically conjugate quantities in phase-space \cite{weyl}.   Slightly more tangibly, perhaps, $\hbar$ appears in the periodicity of thermodynamics at temperature $T$ under translations in imaginary time $\tau = \hbar /T$ \cite{imaginary time periodicity}.     

\subsection{Units of Fundamental Constants}

Since we have established a unit of action, the world-action $\int d^4x \ {\cal L_{\rm world}}$ that defines the standard model is a purely numerical quantity.   The kinetic terms 
\begin{eqnarray}\label{fancyKinetic}
{\cal L}_{\phi\ \rm{kinetic}} &=& \frac{1}{2} \sqrt{g } g^{\alpha \beta} \partial_\alpha \phi \partial_\beta \phi \\
{\cal L}_{\psi\  \rm{kinetic}} &=& \frac{1}{2} \sqrt{g } e^\alpha_a \bar \psi \gamma^a \overleftrightarrow{\nabla}_\alpha \psi
\end{eqnarray}
for scalar and spinor fields, respectively, therefore show that the fields $\phi, \psi$ have the units $[L]^{-1}, [L]^{-\frac{3}{2}}$, respectively, where $[L]$ is the unit of $dx$.  (I've written these equations their full general relativistic glory, including the volume factor $\sqrt{g}$, the metric $g^{\alpha \beta}$, the vierbein $ e^\alpha_a$, and the covariant derivative $\nabla$, which includes a spin connection term, just this once to emphasize a point I'll reiterate later, namely that no immediate difficulty or ambiguity arises in incorporating gravity, including its quantum mechanics, within the traditional standard model.   In non-gravitational physics it is usually adequate to use the limiting flat-space forms: $\sqrt{g} =1$, $g^{\alpha \beta} = \eta^{\alpha \beta}$ the signed Kronecker delta $(1, -1, -1, -1)$, $e^\alpha_a$ the ordinary Kronecker delta, and $\nabla \rightarrow \partial$.)   Similarly, gauge potential (vector) fields $A_\alpha$ have units $[L]^{-1!
 }$.  

\subsubsection{Units in the conventional standard model}

Interactions in the standard model involve various products of scalar, spinor, and vector fields appearing as additive contributions to $\cal {L}_{\rm world}$, with coefficients partially but not entirely fixed by symmetry.   Since the units of the field have been fixed, as are the (trivial) units of $\int d^4x\  {\cal L_{\rm world}}$, the units of these coefficients are uniquely fixed, as various powers of $[L]$.    

Thus all the parameters of the standard model can be specified in terms of a single unit.   To do so is of course a very well established practice, but its ultimate logical foundation, explained here, seems to be stated rarely (if ever).   The unit can be taken as the unit of length, as above.  In high-energy physics it is usually more convenient to take mass, energy, or momentum as the fundamental unit.  Each of these is equivalent to $[L]^{-1}$, reflecting the relations $[M] = \frac{\hbar}{c} [L]^{-1}$, $[E] =  \hbar c  [L]^{-1}$, $[P] = \hbar  [L]^{-1}$.   

For the following discussion it will be convenient to define the mass dimension of a quantity to be the power of $[M]$ that appears in its unit.   Thus scalar and vector fields have mass dimension 1, and spinor fields have mass dimension $\frac{3}{2}$.

Local interaction terms
are obtained from Lagrangian densities involving products of fields
and their derivatives at a point.   The coefficient of such a term is
a coupling constant, and must have the appropriate mass dimension to insure
that each term in  the Lagrangian density has mass dimension 4.

The gauge couplings of the standard model are implemented by minimal coupling procedure, promoting ordinary to covariant derivatives:
\begin{equation}
\partial_\alpha \rightarrow \partial_\alpha + i \sum_j g_j \tau_j^a A^a_\alpha 
\end{equation}
where the $\tau^a$ are appropriate numerical matrices representing the Lie algebra of each symmetry group. (For the abelian hypercharge group, they are simply real numbers.)   Since both $\partial_\alpha$ and $A^a_\alpha$ have mass dimension one, consistency requires that  the gauge couplings $g_j$ have mass dimension zero; i.e. that the gauge couplings are (dimensionless) pure numbers.    

The generalized masses of quarks and charged leptons (``generalized'', as we'll see, to include weak mixing angles), arise from Yukawa terms in the Lagrangian, of the general form
\begin{equation}
{\cal L}_{\rm Yukawa} = y \bar{\psi} \phi \psi
\end{equation}
where $\psi$ is a spinor fermion field and $\phi$ is a scalar field (the Higgs field).   Since the total mass dimension must be four, and the mass dimensions of $\psi, \phi$ are $\frac{3}{2}, 1$ respectively, we see that Yukawa couplings such as $y$ are likewise dimensionless.   

The potential of the Higgs field
\begin{equation}
{\cal L}_{\rm \phi\ potential} = \mu^2 |\phi |^2 - \zeta |\phi |^4
\end{equation}
brings in yet another dimensionless parameter $\zeta$ and a parameter $\mu$ with mass dimension one.

These are all the coupling types that appear in the conventional standard model (i.e., excluding gravity and neutrino masses).   Thus all the fundamental constants are dimensionless, except for $\mu^2$, which has mass dimension 2.   

\subsubsection{Renormalizability}

The fact that the strong and electroweak interactions, together with the generalized masses of quarks and charged leptons, can be described using only fundamental parameters whose units are non-negative powers of mass is profound.  Indeed, it is difficult to accommodate {\it fundamental\/} constants whose units are negative powers of mass in quantum field theory, for the following reason.   Consider the effect of treating a given
interaction term as a perturbation.  If the coupling $\kappa$
associated to this interaction has negative mass dimension $-p$, then
successive powers of it will occur in the form of powers of 
$\kappa \Lambda^p$,
where $\Lambda$ is some parameter with dimensions of mass.  As explained in the following paragraph,  the interactions in a local field theory are {\it hard}, in the sense that the bring in couplings to arbitrarily high frequency modes, with no suppression.    Thus we
can anticipate that $\Lambda$ will characterize the largest mass scale
we allow to occur (the cutoff),   and will diverge to infinity as the
limit on this mass scale is removed.  So we expect that it will be
difficult to make sense of fundamental interactions having negative
mass dimensions, which bring in powers of $\kappa \Lambda^p$, at least in perturbation theory.  Such interactions
are said to be nonrenormalizable. 

Now we return, as promised, to the concept of ``hardness''.  In order to construct the local field $\psi (x)$ at a
space-time point $x$, one must take a superposition
\begin{equation}
\psi(x) ~=~ \int {\frac{d^4k}{(2\pi )^4}} e^{ikx} \tilde
\psi (k)
\end{equation}
that includes field components $\tilde \psi (k)$
extending to arbitrarily large momenta.  Moreover in a generic
interaction
\begin{equation}
\int {\cal L} = \int \psi (x)^3 = \int
{d^4k_1\over (2\pi )^4}{d^4k_2\over (2\pi )^4}{d^4k_3 \over (2\pi )^4}
\tilde \psi (k_1) \tilde \psi (k_2) \tilde \psi (k_3) (2\pi
)^4\delta^4(k_1 + k_2 +k_3)
\end{equation}
we see that a low momentum
mode $k_1 \approx 0 $ will couple without any suppression factor to
high-momentum modes $k_2$ and $k_3 \approx -k_2$.  Local couplings are
``hard'', in this sense. Because locality requires the existence
of infinitely many degrees of freedom at large momenta, with unsuppressed 
interactions, ultraviolet divergences of the type anticipated in the preceding paragraph actually do occur.  

Thus nonrenormalizable interactions, if extrapolated up to arbitrarily high energy-momenta, become problematic.
We get an extremely precise, and accurate, account of the strong and electroweak interactions using just the coupling types (minimal gauge, Yukawa, Higgs potential) mentioned above, not allowing nonrenormalizable interactions.  

Outstanding examples of this  precision and accuracy are the comparison of electron and muon magnetic moments to measurements, where the agreement extends to parts per billion or better \cite{magnetic moments}.    The electron and muon magnetic moments are corrected from their classical values by contributions from quantum fluctuations: that is to say contributions from loop graphs, or from interactions with virtual particles.   Calculation of those contributions involves exquisite use of the detailed algorithms of quantum field theory, carried to high order in the interactions.  Thus the agreement provides impressive evidence that these algorithms, applied to the Lagrangian of the standard model, correctly describe Nature.   

Yet neither the symmetries of the standard model nor the principle of locality forbid one to include an {\it intrinsic \/} magnetic moment interaction of the type
\begin{equation}
{\cal L}_{\rm moment} ~=~ \kappa \bar e \sigma^{\mu \nu} e F_{\mu \nu}
\end{equation}
Such a term will destroy the agreement of theory and experiment unless the coefficient $\kappa$ is very small.    The mass dimension of $\kappa$ is -1, and it is constrained to be $\lesssim (10\ {\rm TeV})^{-1}$.    
Similarly, the successful comparison of measured weak interaction processes with predictions from the CKM framework \cite{CKM} would be ruined by the presence of significant nonrenormalizable interactions of the general four-fermion type
\begin{equation}
{\cal L}_{\rm 4\ fermion} ~=~ \eta^{ij}_{kl}  \bar \psi_i \psi^k \bar \psi_j \psi^l
\end{equation}
where the mass dimension of $\eta$ is -2.   

Four-fermion interactions were the basis of an older theory of the weak interactions, with roots in Fermi's theory of $\beta$ decay, later generalized into the $V-A$ current-current theory.  From today's perspective, we recognize the older theory as an effective low-energy description governing interactions at energy-momenta well below the masses of the $W$ and $Z$ bosons.  Indeed, the older theory -- extended to include neutral currents --  arises as an approximation to the standard model, obtained by ``integrating out'' the $W$ and $Z$ bosons.   In terms of Feynman graphs, we replace propagator denominators with their low-energy limit:
\begin{equation}
\frac{1}{p^2 - M^2} ~\rightarrow~ \frac{1}{-M^2}
\end{equation}
Thus nonrenormalizable interactions can appear in effective theories, with coefficients that reflect the scale of their more fundamental origin.   

From this perspective, it is plausible that the smallness of contributions from nonrenormalizable interactions can be interpreted as follows:
\begin{itemize}
\item The standard model is not complete, but it is embedded in a larger theory with good high-energy behavior.
\item There is a significant separation of scales, so that the factors $\frac{1}{M^p}$ that arise from integrating out heavy modes within the larger theory are very small (i.e., $M$ is large).
\end{itemize}
Further evidence for this viewpoint emerges from the theory of neutrino masses and of gauge coupling unification, as we'll discuss momentarily.   

Whatever its justification, the assumption that nonrenormalizable interactions can be neglected is a powerful guiding principle, because the remaining possibilities for couplings, with non-negative mass  dimension, are very restricted.   The coupling types mentioned above basically exhaust the possibilities.  Conversely, all the renormalizable interactions
consistent with the gauge symmetry and multiplet structure of the conventional 
standard model do seem to occur -- ``what is not forbidden, is
mandatory''.   There is a beautiful agreement between the symmetries
of the standard model, allowing arbitrary renormalizable interactions,
and the symmetries of the world.  One understands on this basis, for example, why strangeness is observed to be violated,
while baryon number is not.  

The only discordant element  is
the so-called $\theta$ term of QCD, which is allowed by the symmetries
of the standard model but is measured to be quite accurately zero.   
A plausible solution to this problem exists.  It involves a
characteristic very light {\it axion\/} field, as we shall discuss below.

\subsubsection{Neutrino masses}

To obtain non-zero values of neutrino masses using the degrees of freedom available in the standard model, we must allow nonrenormalizable interactions.   A conventional mass term $\propto \bar \nu \nu$ is not possible, because the neutrino field $\nu$ is left-handed.   A so-called Majorana mass term, of the form 
\begin{equation}
{\cal L}_{\rm Majorana\ mass\  plain} ~\propto~ \epsilon_{ij} \nu^i \nu^j
\end{equation}
where we write the left-handed neutrino field in two-component form, is kinematically allowed.  However, 
because the neutrino fields have $SU(2)\times U(1)$ quantum numbers $(\frac{1}{2},  -\frac{1}{2})$, a fundamental term of this form necessarily violates $SU(2)\times U(1)$.    Since the minimal Higgs field $\phi$ also has $SU(2)\times U(1)$ quantum numbers $(\frac{1}{2},  -\frac{1}{2})$ quantum numbers, a term 
\begin{equation}
{\cal L}_{\rm Majorana\ mass\ symmetric} ~=~ \eta^{ab} \epsilon_{ij} L^{i\alpha}_{a} L^{j\beta}_{b} \phi^\dagger_\alpha \phi^\dagger_\beta
\end{equation}
is allowed.  Here $a, b$ are flavor indices and $\alpha, \beta$ are weak $SU(2)$ indices.  (Neutrinos are in doublets with charged leptons, so for $\alpha =1$ $L^\alpha$ is a neutrino field, while for  $\alpha =2$ $L^\alpha$ is a charged lepton field).   When $\phi^1$ acquires a non-zero vacuum expectation value, ${\cal L}_{\rm Majorana\ mass\ symmetric}$ induces a matrix of ${\cal L}_{\rm Majorana\ mass\ plain}$ terms.    This matrix, together of course with the kinetic term, describes the propagation of neutrinos.   Its off-diagonal entries describe neutrino oscillations.   
The measured values of neutrino masses, $\lesssim 10^{-2}\ {\rm eV}$, indicate that the scale for $\eta^{ab}$ is $\sim (10^{16} \ {\rm GeV})^{-1}$, and the pattern of observed oscillations indicates that its structure is complicated.    

Thus the fact that neutrino masses are so small compared to other fermion masses is tied to the more general phenomenon that nonrenormalizable interactions are suppressed.  Following our analogy with the old weak interaction theory, the {\it small \/} value of neutrino masses betokens a very {\it large\/} hidden mass scale.    As we proceed we'll see the same hidden scale appearing twice more, in ways that on the surface appear very different.  Their ``coincidence'' suggests the possibility of  a major synthesis.  

It is appropriate to add three comments:
\begin{itemize}
\item Effective neutrino mass terms, of the kind just discussed, are the only consistent (local, gauge invariant) terms we can construct in the conventional standard model -- using its known degrees of freedom -- whose coefficient has mass dimension -1.   
\item Consistent four-fermion interactions with coefficients of order $M^{-2}$, $M \sim 10^{16}\ {\rm GeV}$, with one class of exceptions, would be too small to be observed.   Thus they could be present without our being aware of it.   The exceptions are interactions that mediate baryon number violation.    For those, $M \sim 10^{16}\ {\rm GeV}$ is near the edge of existing experimental limits.   
\item Though the traditional criterion of renormalizability must be relaxed to accommodate neutrino masses within the conventional standard model, a simple generalization provides a useful guiding principle.   That is, we look to accommodate the new phenomenon with interaction terms of the lowest possible mass dimension -- and thus with coefficients of the highest possible mass dimension.   That principle leads us to the specific and reasonably tight framework sketched above -- which, so far, has proved adequate.   
\end{itemize}

\subsubsection{Gravity} 

Our best working theory of gravity is general relativity.  As was sketched in the discussion following Equation (\ref{fancyKinetic}), it is straightforward to couple the characteristic metric field of general relativity to matter, using a minimal coupling procedure similar to that we employ for gauge fields.   This procedure is motivated, in view of the preceding discussion, by the same principle of keeping interactions terms with the lowest possible mass dimension.   However the Einstein-Hilbert term
\begin{equation}
{\cal L}_{\rm Einstein-Hilbert} ~=~ \frac{1}{16\pi G} \int d^4x \sqrt g \  R
\end{equation}
that governs graviton propagation, is quite different from a conventional kinetic term.  (Here $G$ is the Newtonian gravitational constant and $R$ is the Ricci curvature.)    Only if we expand $g_{\alpha \beta}$ around flat space in the form
 \begin{equation}
g_{\alpha \beta} =
\eta_{\alpha\beta} +  \sqrt G h_{\alpha\beta}
\end{equation}
do we define a conventionally normalized boson field $h_{\alpha\beta}$ of mass dimension 1 (for then the $G$ cancels).   Thus only with this choice will we obtain propagators of the usual form.   This ``renormalization'' of the field $h$ means that its couplings to matter are accompanied by a factor $(\sqrt G)$.   

By the now-familiar dimensional analysis of units, we find that $G$ has mass dimension -2.   Its magnitude is by definition $M_{\rm Planck}^{-2}$, where $M_{\rm Planck} \sim 10^{18}\  {\rm GeV }$ is another fundamental constant, the famous Planck mass.   Note that the Planck mass does not greatly differ from the large mass scale we inferred from neutrino masses.

Thus the couplings of gravitons to matter have negative mass dimension: they are nonrenormalizable.   So too are the nonlinear self-interactions of $h$.   One expects divergences in perturbation theory, and indeed one finds them.   Nevertheless, if one works to lowest in perturbation theory, not including gravitons in loops, one obtains an excellent theory of gravity.   It is, of course, the theory implicitly assumed by practicing physicists and astrophysicists in their everyday work.   It accurately describes all the classic applications of Newtonian and Einsteinian gravity, from precession of the equinoxes to binary pulsar spin-down and gravity waves.   It is also consistent with quantum kinematics, in the sense that the uncertainty relation is obeyed, and gravitons appear as quanta of the gravitational field.  The quantum behavior of matter in gravitational fields is described precisely and accurately by this theory, as attested by many terrestrial and astrophysical measurements (corrections to GPS, red shift of spectral lines, etc.).  

Since loops are divergent, one cannot use this theory of gravity to calculate radiative corrections, just as one could not use the old Fermi theory of weak interactions for that purpose.    However the gravitational radiative corrections are expected to be suppressed by positive powers of ${p}{\sqrt G} \approx \frac {p}{M_{\rm Planck}}$, where $p$ is a characteristic energy-momentum of the process under consideration, and thus to be very small in practice.     This expectation is consistent with all existing observations, which agree with the minimal theory.   

Of course, it would be very desirable to have a complete, logically consistent theory of quantum gravity that could be extrapolated up to arbitrarily high energy-momenta -- or to demonstrate convincingly that those concepts break down!   Such a theory might give new phenomena or unexpected relations among known phenomena, similarly to how the passing from the Fermi theory to the modern theory of electroweak interactions led us to predict the existence of neutral currents, to the CKM framework, and to an attractive framework for accommodating neutrino masses.   Such a theory might also allow us new insight into situations of extreme curvature (large $p$!) such as might occur inside black holes or in the earliest stages of the Big Bang.   It is far from true, however, that the absence of a complete theory of quantum gravity puts physics into crisis along a broad front.  On the contrary, it is challenging to identify specific phenomena for which the standard model, as defined h!
 ere to include gravity, might be subject to measurable corrections from a more complete theory of quantum gravity.

Finally, the simplest of all interactions consistent with the principles of the standard model is 
\begin{equation}
{\cal L}_{\rm dark\ energy} ~=~ - \lambda \int d^4x \sqrt g 
\end{equation}
Such a term leads to Einstein's cosmological term, now often referred to as ``dark energy''.   This term could arise, for example, as an offset to the zero of the Higgs potential.   In the cosmological equations, it provides (with $\lambda >0$) a source of positive density $\rho_\lambda$ and negative pressure $p_\lambda = -\rho_\lambda$.    Observations indicate that $\rho_\lambda$ provides about 70\% of the mass of the universe as a whole, and that $p_\lambda$ is beginning to cause the expansion of the universe to {\it accelerate}. 

$\lambda$ has mass dimension 4 -- thus it is {\it very\/} renormalizable.  The astronomical observations correspond to $\lambda \approx (10^{-3}\ {\rm eV} )^{4}$.   There are two conceptual difficulties with this value: that it is so small, and that it is so large.

A great lesson of the standard model is that what human senses have been evolved to
perceive as empty space is in fact a richly structured medium. ``Empty'' space contains
symmetry-breaking condensates associated with electroweak
superconductivity and with spontaneous chiral symmetry breaking in QCD, an
effervescence of virtual particles, and probably much more. Since
gravitons are sensitive to all forms of energy they really ought to see this stuff,
even if we don't.  Straightforward estimation suggests that empty space
should weigh several orders of magnitude of orders of magnitude
(no misprint here!) more than it does.  It ``should" be much denser than a
neutron star, for example. The expected energy of empty space acts like
dark energy, with negative pressure, but far more is expected than is observed.  Given this discrepancy, many physicists hoped that some new principle would emerge -- perhaps a consistency requirement from quantum gravity -- that would constrain $\lambda$ to vanish.   Evidently those hopes, at least in their simplest form, have been dashed.

Speculative ideas \cite{uniOrMulti} aiming to explain the observed value of $\lambda$ are discussed at length elsewhere in this volume, and briefly below.

%%%%%%%%%%%%%%%%%%%%

\subsection{Closed Unit Systems}

It is instructive and entertaining to connect the preceding discussion of units and fundamental constants with others that have appeared in the literature, or are natural to consider.  

As we've seen -- or rather, perhaps more accurately, as I've tried to argue -- it is natural to take the standard model as the basis for defining units and fundamental constants.  And in the context of the standard model it is natural to regard $c$ and $\hbar$ as the units of velocity and action.   We then find that all the fundamental constants can be defined in terms of a single unit, which we can take to be a unit of mass (or energy or momentum or inverse length).   To complete the system of units, we should add one more dimensional quantity.    Within the standard model, a natural choice would appear to be $\mu$.    On the other hand, we won't know its value until the Higgs particle is discovered; and we might well find that the Higgs sector is non-minimal, in which case no simple parameter $\mu$ exists as such.   

With that in mind, let us consider some important alternatives. 
\begin{itemize} 
\item Planck introduced \cite{Planck} a famous system of units at the dawn of quantum theory.   He implicitly assumed that three units are required, namely the mechanical units $[M], [L], [T]$.   Planck stressed that it is possible to construct such units from the universal parameters $\hbar, c, G$, in the form 
\begin{equation}
\label{planckUnits}
[M]_{\rm Planck} = \sqrt {\frac {\hbar c}{G}},\  [L]_{\rm Planck} = \sqrt {\frac {\hbar G}{c^3}}, \ [T]_{\rm Planck} = \sqrt {\frac {\hbar G}{c^5}}
\end{equation}
These units are called, of course, Planck units.   In general relativity, we learn that energy-momentum causes space-time curvature.  But these quantities are measured in different units, so we need a conversion factor.   $G$ supplies that conversion factor.  Since $G$ appears in such a central role in such a profound phenomenon of physics, it is natural to think that it will appear as a primary ingredient in the formulation of a complete, unified theory of physics.   Thus we might anticipate, according to the usual assumption of dimensional analysis, that in such a theory  all fundamental quantities will appear, in Planck units, as pure numbers of order unity.   That program is challenging, because most masses of elementary particles are actually extremely tiny in Planck units.    The smallness of the proton mass, $m_p \sim 10^{-18} [M]_{\rm Planck}$, has a profound interpretation, as I'll indicate below.   The smallness of the Higgs mass parameter $\mu \sim 10^{-16} [M]_{\!
 rm Planck}$ is known as the ``hierarchy problem''.   
\item Prior to Planck, Stoney introduced \cite{Stoney} units based on $e, c, G$.   Algebraically this is not very different from Planck's system: since the fine structure constant $\alpha \equiv \frac{e^2}{4\pi \hbar c}$ is a pure number $\sim \frac{1}{137}$ , one can simply trade $\hbar$ for $\frac{e^2}{c}$.    In this system of units, $\hbar$ is a derived quantity.    Thus to build up a complete fundamental theory in terms of pure numbers and $e,c,G$, one would need to find quantum mechanics as an emergent phenomenon.   On the positive side, it is interesting that after the substitution $\hbar \rightarrow \frac{e^2}{ c}$ $e$ and $c$ can be taken outside the square roots that appear in Eqn. (\ref{planckUnits}) for $[M],[L],[T]$.   
\item Atomic units are based on $e, \hbar, m_e$, where $m_e$ is the mass of the electron.   The nonrelativistic Schr\"odinger equation, with nuclei idealized as infinitely massive point sources of charge, becomes dimensionless in these units.   Thus the sizes and shapes of molecules become, within that approximation, purely numerical quantities.   $m_e$ does not appear directly in our list of fundamental constants; it is a slightly complicated derived quantity, as we'll discuss further below.   So this useful system, that expresses a profound truth of structural chemistry, may prove awkward for deeper levels of reduction.
\item Strong units are based on $\hbar, c, m_p$.   This is quite a different completion of the standard model units from Planck's.   Strong units are obviously convenient for work in QCD and nuclear physics, where quantum mechanics and relativity are omnipresent and the proton is an object of central interest.   And $m_p$ can be very precisely measured.  On the other hand, $m_p$ is {\it not\/} a fundamental constant in the sense defined here.  In fact the proton is, in terms of fundamental quarks and gluons, quite a complex object, and its mass $m_p$ is a very complicated derived quantity.    More closely related to fundamentals is the mass parameter $\Lambda_{\rm QCD}$, which parameterizes the behavior of the energy-dependent strong coupling ``constant''.   (More on this below.)   Unfortunately $\Lambda_{\rm QCD}$ is complicated to define and hard to measure precisely.  Since $m_p$ is both  closely related to $\Lambda_{\rm QCD}$ conceptually and not grossly different in value, $m_p$ seems a better practical choice.   In the strong system of units no square roots at all appear in  $[M],[L],[T]$.   
\end{itemize}

The philosophical significance of a complete set of units, is that it allows us to express any fundamental constant as a pure number.  According to the ideal of theoretical physics expressed by Einstein
\begin{quote}
I would like to state a theorem which at present can not be based upon anything more than upon a faith in the simplicity, i.e., intelligibility, of nature: there are no arbitrary constants ... that is to say, nature is so constituted that it is possible logically to lay down such strongly determined laws that within these laws only rationally completely determined constants occur (not constants, therefore, whose numerical value could be changed without destroying the theory).
\end{quote}
we must aspire to calculate all those numbers.

\section{Four Kinds of Fundamental Constants}

It is a remarkable fact, that
every nonlinear
interaction we need to summarize our present knowledge of the basic 
laws of physics involves one of three kinds of particles: gravitons, vector gauge particles, or Higgs particles.  We've already encountered these couplings in the preceding section; here we'll bring out the geometric interpretation of the gauge and gravitational couplings, and contrast those elegant structures with the accommodation of inertia in the Higgs sector.  

\subsection{Internal Curvature}

To bring out the geometric nature of the gauge couplings, it is convenient to use slightly rescaled fields, absorbing the coupling constants:
\begin{eqnarray}
\tilde A^a_\alpha ~&\equiv&~ g A^a_\alpha \\
\tilde F^{a}_{\ \alpha \beta}  ~&\equiv&~ \partial_\alpha \tilde A^a_\beta - \partial_\beta \tilde A^a_\alpha + f^{abc} \tilde A^a_\alpha \tilde A^b_\beta
\end{eqnarray}
where the $f^{abc}$ are the group structure constants.   (For simplicity, indices distinguishing the different gauge groups have been suppressed.) 
Then the coupling constants disappear from the definition of covariant derivatives.   They appear only as the coefficients of the gauge kinetic terms, in the form
\begin{equation}
{\cal L}_{\rm gauge\ kinetic} ~=~ -\frac{1}{4g^2} \tilde F^a_{\ \alpha \beta} \tilde F^{a\alpha \beta}
\end{equation}

The normalization of the kinetic terms is canonical for $A$, so it becomes non-standard in terms of $\tilde A$.   In terms of Feynman graphs, the propagators for $\tilde A$ will contain factors $g^2$, but the vertices will be free of coupling constants.   Thus the {\it universality\/} of the coupling is manifest.  (In the nonabelian case, the only choice in couplings is a discrete choice: the choice of a representation.   In the abelian case, the numerical bare charges can still be chosen arbitrarily; in that case, universality entails that the ratio of physical charges is same as the ratio of bare charges, a statement usually formalized as Ward's identity.)  We encountered a very similar situation with gravity: the natural formulation of general relativity builds in universality, but involves a non-canonical kinetic energy term.  

The gauge field strength measures the non-commutativity of gauge covariant derivatives:
\begin{equation}
([ \nabla_\alpha , \nabla_\beta ] \phi)^j ~=~ i \tilde F^a_{\ \alpha \beta} \tau^{aj}_k \phi^k
\end{equation}
where $\phi$ is a field in the representation given by $\tau$.   It is therefore analogous to the Riemann curvature of space-time, which measures the failure of space-time covariant derivatives to commute, {\it e.g}.
\begin{equation}
([ \nabla_\alpha , \nabla_\beta ] v)^\gamma ~=~ R_{\ \delta\alpha\beta}^\gamma v^\delta
\end{equation}
for a vector field $v$.  The gauge field strength measures the curvature of the internal spaces. one over each space-time point, in which charged fields such as $\phi^j$ propagate, rotating their indices.   Mathematicians have found it fruitful to take such internal spaces literally, in the theory of fiber bundles and characteristic classes \cite{nakahara}.    In some forms of Kaluza-Klein theory, the internal spaces arise as compactifications of additional spatial dimensions \cite{Kaluza-Klein}.

On this interpretation, the kinetic term $\propto | F |^2$ measures resistance to curvature.  When the coupling is small, curvature comes at a high price in action.  The gauge field is stiff, and does not want to move off zero (or gauge equivalents).   When the coupling is large curvature is cheap, and the gauge field fluctuates freely.  

Another invariant term of mass dimension 4 can be constructed from the gauge curvatures:
\begin{equation}
{\cal L}_{\rm theta} ~=~ \frac{\theta}{16\pi^2}  \tilde F^a_{\ \alpha \beta} \tilde F^{a\gamma \delta} \epsilon^{\alpha \beta \gamma \delta}
\end{equation}
It has remarkable properties and is closely connected to important QCD ``instanton'' physics \cite{coleman}.   From the perspective of fundamental constants, the most important observation is that the interaction associated with a non-zero value of $\theta$ would introduce P and T violation into the strong interaction, which is not found.    Experiments put strong bounds on $\theta$: $|\theta | \lesssim 10^{-9}$.   Theoretical attempts to understand the smallness of $\theta$ lead us to the physics of axions -- of which more below.

\subsection{Space-Time Curvature and Space-Time Volume}

The Einstein-Hilbert term measures the stiffness of space-time, its resistance to curvature.   Unlike the situation for gauge curvature, however, not all forms of space-time curvature carry a high price in action.   Indeed, the vanishing of the positive-definite expression  $\tilde F^a_{\ \alpha \beta} \tilde F^{a\alpha \beta}$ entails $\tilde F^{a\alpha \beta}\ = \ 0$ and thus the triviality of the gauge field, but $\sqrt g R$ is not even positive definite.  Moreover, its variation
\begin{equation}
\delta \sqrt g R ~ = ~ \sqrt g (R_{\alpha \beta} - \frac{1}{2} g_{\alpha \beta} R) \delta g^{\alpha \beta}
\end{equation}
can vanish without the total curvature $R_{\alpha \beta \gamma \delta}$ vanishing.  The non-positivity of the Einstein-Hilbert action raises difficult issues, over and above the problem of nonrenormalizability, for formulating a full-fledged quantum theory based on general relativity.  For example, it is very difficult to see how a well-behaved path integral could emerge.   

The cosmological term ${\cal L}_{\rm dark\ energy} ~=~ - \lambda \int d^4x \sqrt g $ is proportional to the space-time volume.  For $\lambda > 0$, it assigns low action for large space-time volumes -- that is, it makes a large {\it negative\/} contribution to their action.

%%%%%%%%%%%%%%%%%%

\subsection{Accommodations of Inertia}

Thus far in this section we've discussed discussed five fundamental constants: three associated with gauge field curvature, one associated with space-time curvature, and one associated with space-time volume.   These constants have appealing geometric interpretations.  The first four accurately describe an enormous range of phenomena and have been tested in great detail; the last, the cosmological term, describes a few enormously important phenomena.   

The remaining fundamental constants appear as coefficients of the
Higgs field mass and 
self-coupling, and various Yukawa couplings.  All these terms
involve the Higgs field, in one way or another.    None has been measured directly!   

The quantities we've actually measured arise as follows.   (For concreteness I'll focus on the quark sector.  A parallel story holds for the lepton sector, with neutrino masses and mixings arising from slightly more complicated couplings as indicated previously.)   One has two complex matrices $h_a^b, k_a^b$ of couplings, with the indices running over 1, 2, 3, corresponding to the three families.  They appear in the Lagrangian terms
\begin{eqnarray}\label{couplingMatrices}
{\cal L}_{\rm up\ couplings} ~&\equiv&~ - h_a^b \bar {Q_L}_{\alpha b} U_R^a  \phi^\alpha + ({\rm hermitian\ conjugate})  \\
{\cal L}_{\rm down\ couplings} ~&\equiv&~ k_a^b \bar {Q_L}_{\alpha b} D_R^a  \phi^\dagger_\beta \epsilon^{\alpha \beta} + ({\rm hermitian\ conjugate}) 
\end{eqnarray}
Here the Greek indices run from 1 to 2, for vectors of weak $SU(2)$, 
$$
Q^b_L ~\equiv~ \left(\begin{array}{c} U^b_L \\ D^b_L \end{array}\right)
$$
are the left-handed quark doublets, and of course $\phi^\alpha$ is the Higgs doublet.   When $\phi^1$ acquires its vacuum expectation value $v$, breaking electroweak $SU(2)\times U(1)$ down the the $U(1)$ of electromagnetism, the coupling matrices of Equation (\ref{couplingMatrices}) induce the mass matrices
\begin{eqnarray}\label{massMatrices}
{\cal L}_{\rm up\ masses} ~&\equiv&~ - h_a^b v \bar {U_L}_{\alpha b} U_R^a  + ({\rm hermitian\ conjugate})  \\
{\cal L}_{\rm down\ masses} ~&\equiv&~ - k_a^b v \bar {D_L}_{\alpha b} D_R^a  + ({\rm hermitian\ conjugate}) 
\end{eqnarray}
Now to obtain particles with normal propagation properties (i.e., eigenstates of the free Lagrangian) we must make unitary rotations so as to diagonalize these matrices.   Naming the unitary rotation matrices $S_{UL}, S_{UR}, S_{DL}, S_{DR}$, and defining $\tilde M_{Ua}^b \equiv vh_a^b, \tilde M_{Da}^b \equiv vk_a^b$, we require
\begin{eqnarray}
S_{UL}^\dagger \tilde M_U S_{UR} ~&=&~ M_U \\
 S_{DL}^\dagger \tilde M_D S_{DR} ~&=&~ M_D
 \end{eqnarray}
 with $M_U$, $M_D$ positive and diagonal.   
 
 The entries of $M_U$ and $M_D$ are the observable masses of the up-type (charge $\frac{2}{3}$) and down-type (charge $-\frac{1}{3}$) quarks respectively.  The CKM matrix is given as $S_{UL}^\dagger S_{DL}$.    The CKM matrix entries give the weak mixing angles, i.e. family-dependent multiplicative factors in the charged current to which the $W$ boson couples.  (Strictly speaking, $S_{UL}^\dagger S_{DL}$ must be tweaked a little further, to remove some redundant phase factors, before it becomes the CKM matrix.)     
 I've entered into painful detail here to emphasize that the measured masses and mixing angles are rather complicated combinations of fundamental constants.  The real situation could well become even more complicated, if there are several Higgs fields that contribute to quark and lepton masses.

In any case, we know of no deep
principles, comparable to
gauge symmetry or general covariance, that give powerful constraints upon, or relations among, the values of
these couplings.
As a consequence, in this sector the number of continuous fundamental constants increases into the
dozens.  Each observed mass and weak mixing angle is an
independent input, determined empirically, that expresses a complicated combination of fundamental constants, in the way just described.   The literature contains many semi-phenomenological proposals for constraining the choices by imposing various sorts of symmetry; for a fully worked-out example, with many references, see \cite{babu}.

The flavor/Higgs sector of the standard model is, by a wide margin, its least satisfactory part.
Whether judged by the large number of independent parameters or by the small
number of powerful ideas it contains, our theory of
this sector does not
attain the same level as we've reached in the other sectors.  This
part truly
deserves to be called a ``model'' rather than a ``theory''.

\section{Unification of the Curvature Couplings}

\subsection{Unification Though Symmetry Enhancement}

The structure of the gauge sector of the standard model gives
powerful suggestions for its further development. The product
structure $SU(3)\times SU(2) \times U(1)$, the reducibility of the fermion
representation, and the peculiar values of the hypercharge
assignments all suggest the possibility of a larger symmetry,
that would encompass the
three factors, unite the representations, and fix the hypercharges.
The devil is in the details, and it is not at all automatic that the observed,
complex pattern of matter will fit neatly into a simple mathematical structure.
But, to a remarkable extent, it does.  The smallest simple group into
which $SU(3)\times SU(2) \times U(1)$ could possibly fit, that is $SU(5)$,
fits all the fermions of a single family into two representations
($\bf{10} +\bar{\bf 5}$), and the hypercharges click into place. A
larger symmetry
group, $SO(10)$, fits these and one additional $SU(3)\times SU(2) \times
U(1)$ singlet particle into a single representation, the spinor $\bf{16}$.  
The additional particle is actually quite welcome.  It has the quantum
numbers of a right-handed neutrino, and it plays a crucial role in
the attractive ``seesaw" model of neutrino masses.  (See below, and for a more extended introduction to these topics \cite{spacePart}. )

\subsection{Quantitative Unification}

The unification of quantum numbers, though attractive, remains
purely formal until it is embedded in a physical model.  That
requires realizing the enhanced symmetry in a local gauge theory. But
nonabelian
gauge symmetry requires universality: it requires that the relative
strengths of the different couplings must be equal, which is not what
is
observed.

Fortunately, there is a compelling way to save the situation. If the higher
symmetry is broken at a large energy scale (equivalently, a small
distance scale),
then we observe interactions at smaller energies (larger
distances) whose intrinsic strength has been affected by the physics of
vacuum polarization.  The running of couplings is an effect that can be
calculated rather precisely, in favorable cases (basically, for weak coupling),
given a definite hypothesis about the particle spectrum. In this way
we can test, quantitatively, the idea that the observed couplings derive
from a single unified value.

Results from these calculations are remarkably encouraging.
If we include vacuum polarization from the particles we know about in
the minimal standard model, we find approximate unification \cite{GQW}. If we include
vacuum polarization from the particles needed to expand the standard
model to include supersymmetry, softly broken at the TeV scale, we find
accurate unification \cite{susyRunning}.   Within this circle of ideas, called ``low-energy supersymmetry'', we predict the existence of a whole new world of particles with masses in the TeV range.  There must be supersymmetric partners of all the presently known particles, each having the same quantum numbers as known analogue but differing in spin by $\frac{1}{2}$, and of course with different mass.    Thus there are spin-$\frac{1}{2}$ gauginos, including gluino partners of QCD's color gluons and wino, zino, and photino partners of $W,Z,\gamma$, spin-0 squarks and sleptons, and more (Higgsinos, gravitinos, axinos).    Some of these particles ought to become accessible with as the Large Hadron Collider (LHC) comes into operation.

The unification occurs at a very large energy scale $M_{\rm unification}$, of
order $10^{16}$ GeV .  This success is robust against small changes in the
SUSY breaking scale, and is not adversely affected by incorporation of
additional particle multiplets, so long as they form complete representations
of $SU(5)$.

On the other hand, many proposals for physics beyond the standard model
at the Tev scale (Technicolor models, large extra dimension scenarios, most
brane-world scenarios) corrupt the foundations of the unification of
couplings calculation, and would render its success accidental. 

\subsection{Importance of the Emergent Scale}

Running of the couplings allows us to infer, based entirely on low-energy data, an enormously large new mass scale, the scale at which unification occurs.    The disparity of scales
arises from the slow
(logarithmic) running of inverse couplings, which implies that
modest differences in observed couplings must be made up by a long interval of
running.  The appearance of a very large mass scale is profound, and welcome on
several grounds:

\begin{itemize}
\item
Earlier we discussed the accommodation of neutrino masses and mixings within the standard model, through use of nonrenormalizable couplings.   With unification, we can realize those couplings as low-energy approximations to more basic couplings that have better high-energy behavior, analogous to the passage from the Fermi theory to modern electroweak theory.   

Indeed, Right-handed neutrinos can have normal, dimension-four Yukawa couplings to the
lepton doublet. In $SO(10)$ such couplings are pretty much mandatory,
since they are related by symmetry to those responsible for
charge-$\frac{2}{3}$ quark masses.  In addition, since right-handed neutrinos are neutral under
$SU(3)\times SU(2)\times U(1)$
they, unlike the fermions of the standard
model, can have a Majorana type self-mass without violating those low-energy
symmetries.   We might expect the self-mass to arise where it is first
allowed, at the scale where $SO(10)$  breaks (or, in other model of unification, its moral equivalent).
Masses of that magnitude remove the right-handed neutrinos from the accessible
spectrum, but they have an important indirect effect. In second-order
perturbation theory the ordinary left-handed neutrinos, through their
ordinary Yukawa couplings, make virtual transitions to their right-handed
relatives and back.  (Alternatively, one substitutes 
\begin{equation}
\frac{1}{p \!\!\! / -M_{\nu_R}} ~\rightarrow ~ \frac{1}{-M_{\nu_R}}
\end{equation}
in the appropriate propagator.)
This generates non-zero masses for the ordinary
neutrinos that are much smaller than the masses of other leptons and quarks.

The masses predicted in this way are broadly consistent with
the tiny observed neutrino masses.  That is, the mass scale associated with the effective nonrenormalizable coupling, that we identified earlier, roughly coincides with the unification scale deduced from coupling constant unification.   Many, though certainly not all, concrete models of $SO(10)$ unification predict $M_{\nu_R} \sim M_{\rm unification}$.   No more than order-of-magnitude success
can be claimed, because relevant details of the models are poorly
determined.

\item
Unification tends to obliterate the distinction between quarks and leptons,
and hence to open up the possibility of proton decay. Heroic experiments
to observe this process have so far come up empty, with limits on partial
lifetimes approaching $10^{34}$ years for some channels. It is very difficult
to assure that these processes are sufficiently suppressed, unless the
unification scale is very large.  Even the high scale indicated by running of
couplings and neutrino masses is barely adequate. Spinning it
positively, experiments to search for proton decay remain a most
important and promising probe into unification physics.
\item
Similarly, it is difficult to avoid the idea that unification, brings in new
connections among the different families.  There are significant
experimental constraints on strangeness-changing neutral currents, lepton
number violation, and other exotic processes that must be suppressed,
and this makes a high scale welcome.
\item
Axion physics requires a high scale of Peccei-Quinn (PQ) symmetry breaking,
in order to implement weakly coupled, ``invisible" axion models.  (See below.)  Existing observations only bound the PQ scale from below, roughly as $M_{\rm PQ} \gtrsim 10^9\ {\rm GeV}$.   Again, a high scale is welcome.  Indeed many, though certainly not all, concrete models of PQ symmetry suggest $M_{\rm PQ} \sim M_{\rm unification}$.    
\item
The unification scale is 
and electroweak interactions with gravity becomes much more plausible.
Newton's constant has dimensions of mass$^{2}$, so it runs even
classically. Or, to put it less technically, because gravity responds directly to energy-momentum,
gravity appears stronger to shorter-wavelength, higher-energy probes. 

Because
gravity starts out extremely feeble compared to other interactions
on laboratory scales, it becomes roughly equipotent with them only at
enormously high scales, comparable to the Planck energy
$\sim 10^{18}\ {\rm GeV}$. This is not so different from $M_{\rm unification}$.   That numerical coincidence might be a fluke; but it's prettier to think that it betokens the descent of all the curvature interactions from a common source.
\end{itemize}

\subsection{Importance of Low-Energy Supersymmetry}

Low-energy supersymmetry is suggested by the quantitative details of coupling constant unification, as just described.  
Low-energy supersymmetry is desirable on several other grounds, as well.

In the absence of supersymmetry radiative
corrections to the vacuum expectation value of the Higgs particle diverge,
and one must fix its value (which, of course, sets the scale for electroweak
symmetry breaking) by hand, as a renormalized parameter. That leaves it
mysterious why the empirical value is so much smaller than unification
scales.

Low-energy supersymmetry protects the Higgs (mass)$^2$ term, which governs
the scale of electroweak symmetry breaking, from quadratically
divergent radiative
corrections.  As long as the scale of mass splittings between
standard model particles
and their superpartners is less than a TeV or so, the radiative
corrections to this
(mass)$^2$ are both finite and reasonably small.  (In detail, things
are not quite so clean
and straightforward; there is the ``$\mu$ problem'', which is a very
interesting and
important subject, but too intricate to discuss here.)

Upon more detailed consideration the challenge takes shape and
sharpens considerably. Enhanced unification symmetry requires that the
Higgs doublet should have partners, to fill out a complete representation.
However these partners have the quantum numbers to mediate proton
decay, and so if they exist at all their masses must be very large, of
order the unification scale $10^{16}$ GeV .  This reinforces the idea
that such a
large mass is what is ``natural" for a scalar field, and that the light doublet
we invoke in the standard model requires some special justification. It
would be facile to claim that low-energy supersymmetry by itself
cleanly resolves these problems, but it does provide powerful
theoretical tools for
addressing them.

The qualitative relationship between mass splittings of supersymmetry
multiplets and the observed weak scale penetrates also has a more specific and
quantitative aspect.  Supersymmetry relates the physical mass of the lightest,
``standard model-like" Higgs particle, which in the absence of
supersymmetry is a free
parameter, to the masses of W and Z bosons.  There is some model
dependence in this
relationship. But within minimal or reasonably economical
supersymmetric extensions
of the standard electroweak model the Higgs mass is generally
predicted to be near -- or
below! -- existing experimental limits \cite{higgsEstimates}.  This
renders the models
subject to quick falsification at LHC or, more optimistically, to
fruitful vindication.

That optimistic scenario gains credibility from another advantage of
supersymmetry.
Supersymmetry has the important though negative virtue, that it yields only small corrections from the standard model predictions for electroweak radiative corrections.
That's a good thing,
because the measurements agree remarkably well with standard model predictions.  

Several large classes of rival models to
low-energy
supersymmetry associate electroweak symmetry breaking with new strong
interactions.  In these models, which include Technicolor and both in
its original form
and in its extra-dimensional disguises, radiative corrections to the
Higgs (mass)$^2$ are
rendered finite by form-factors, rather than cancellations.
Though the additional
radiative contributions in these models are finite, there is no
general reason to expect
that they are especially small.  Indeed, to the extent that they
support specific
calculations, one finds that such models generically have severe difficulty in
accommodating existing precision measurements.

Finally, low-energy supersymmetry can provide an excellent candidate to provide
the dark matter of cosmology.  It's plausible that the lightest
particle with odd
$R$-parity, where $R \equiv  (-)^{3B+L+2J}$ is stable on cosmological
time scales, because
the quantum numbers that go into the definition of $R$ are well
respected.   The lightest
$R$-odd particle, usually called the LSP (Lightest Supersymmetric
Particle) could be
some linear combination of the photino, zino, and Higgsino.  Indeed,
the production of
these particles in big bang cosmology is about right to account for
the observed density
of dark matter.

\section{The Constants of Cosmology}

In recent years a second impressive ``standard model'' has emerged, a standard model of cosmology. 
Like the standard model of fundamental physics, the standard model of cosmology requires us to specify the values of a few parameters.   Given those parameters, the equations of the standard model of cosmology describe important features of the content and large-scale structure of the universe.    

\subsection{Inventory}
 
It is convenient to think of the standard model of cosmology as consisting of two parts.
One part of it is simply a concrete parameterization of the equation of
state to insert into the framework of general relativistic models of
a spatially
uniform expanding Universe (Friedmann-Robertson-Walker model).  The
other part is a very specific hypothesis about the small primordial fluctuations from
uniformity.

Corresponding to the first part, one set of parameters in the
standard model of cosmology specifies a few average properties of matter,
taken over large spatial volumes. These are the densities of ordinary
matter (i.e., of baryons), of neutrinos, of dark matter, and of dark energy.

We know quite a lot about ordinary matter, of course, and we can detect
it at great distances by several methods. It contributes about 5\% of
the total density.  

We have a pretty reliable theory to predict the production of neutrinos during the big bang.  Thus given the magnitude of neutrino masses, we can predict their contribution to the cosmic mass budget.   Unfortunately neutrino oscillations are sensitive only to mass differences, and direct laboratory bounds on the sum of the masses are much looser.   That is, the measured mass differences are considerably smaller -- by roughly two orders of magnitude -- than the bound on the absolute mass.   Cosmology actually provides the best limit on the absolute mass \cite{tegmark}. 

Concerning dark (actually, transparent) matter we know much less.  It
has been ``seen"
only indirectly, through the influence of its gravity on the motion
of visible matter.
We observe that dark matter exerts very little pressure, and
that it contributes about 25\% of the total density.

Finally dark (actually, transparent) energy contributes about 70\% of the
total density.  It has a large {\it negative\/} pressure. From the point
of view of fundamental physics this dark energy is quite mysterious and
disturbing, as mentioned previously.

Given the constraint of spatial flatness, these four densities are
not independent. They must add up to a critical density that
depends only the strength of gravity and the rate of expansion of the universe.

Fortunately, our near-total ignorance concerning the nature of most
of the mass of the Universe does not bar us from modeling the evolution of its density.
That's because the dominant interaction on large scales is gravity, and
gravity does not care about details.
According to general relativity, only total energy-momentum counts -- or
equivalently, for uniform matter, total density and pressure.

Assuming the above-mentioned values for the relative densities, and that the geometry
of space is flat -- and still assuming uniformity -- we can use the equations
of general relativity to extrapolate the present expansion of the Universe
back to earlier times. This procedure defines the standard (uniform) Big Bang scenario.
The Big Bang scenario successfully predicts several things that would otherwise be very
difficult to understand, including the red shift of distant galaxies,
the existence of the microwave background radiation, and the relative abundance
of light nuclear isotopes.  It is also internally consistent, and even
self-validating, in that the microwave background is observed to be uniform to
high accuracy, namely to a few parts in $10^5$.

The other parameter in the standard model of cosmology
concerns the small departures from uniformity in the early Universe.
The seeds grow by gravitational instability, with over-dense regions
attracting more matter, thus increasing their density contrast with time.
Plausibly this process could, starting from very small seeds, eventually
trigger the formation of galaxies, stars, and other structures
we observe today.  {\it A priori\/} one might consider all kinds
of assumptions about
the initial fluctuations, and over the years many hypotheses
have been proposed.  But recent observations, especially the recent,
gorgeous WMAP
measurements \cite{WMAP} of microwave background anisotropies, are broadly consistent with what
in many ways is the simplest possible guess, the so-called Harrison-Zeldovich
spectrum. In this set-up the fluctuations are assumed to be strongly
random -- uncorrelated and Gaussian with a scale invariant spectrum at
horizon entry, to be precise -- and to affect both ordinary and dark
matter equally (adiabatic fluctuations).  Given these strong
assumptions just one
parameter, the overall amplitude of fluctuations, defines the statistical
distribution completely. With an appropriate value for this amplitude, and
the relative density parameters I mentioned before, this standard
model cosmological model fits the WMAP data and other measures of large-scale
structure remarkably well.  

The latest WMAP results may indicate small departures from scale invariance. Other refinements of the basic model may be required in the future, to accommodate departures from Gaussian statistics, separate fluctuations in dark matter and baryon density (so-called isocurvature, as opposed to adiabatic, fluctuations), or primordial gravity waves.   But for the moment we can construct an adequate model universe using equations that contain just four parameters: the four densities, constrained so that their total is the critical density, and the amplitude of the primordial (adiabatic, scale invariant) spectrum.

\subsection{Searching for Foundations}

In the preceding inventory, cosmology has been ``reduced" to some general hypotheses and just
four exogenous parameters: the densities of ordinary baryonic matter, neutrinos, dark matter, and dark energy, constrained to sum up to the critical density, and the amplitude of primordial fluctuations.  Since the neutrino density can be calculated in terms of standard model parameters, as we've discussed, really it's down to three.  
The experimental validation of this cosmological world-model  is an amazing development. Yet I think that most physicists will
not, and should not,
feel entirely satisfied with it.  

For one thing, since the cosmological parameters appear within an open, semi-phenomenological framework, there's every reason to anticipate that as observations improve we'll require more, as we just discussed.

Also, the parameters appearing in
the cosmological model, unlike those in the standard model which they superficially resemble, do not
describe the fundamental behavior of simple entities. Rather they
appear as summary descriptors of averaged properties of macroscopic
(VERY macroscopic!) agglomerations.  The working parameters of cosmology appear neither as key players in
a varied repertoire of phenomena nor as essential elements in a beautiful
mathematical theory.   We'd like to carry the analysis to
another level,
where the four working parameters will give way to different
ones that are, in those ways, closer to fundamentals.   

There's an inspiring model for progress of that kind.  Fifty years ago the relative abundance of each nuclear isotope would have had to be considered an irreducible parameter of cosmology.  Those parameters could not be calculated in terms of anything more basic -- and there are dozens of them.   Now we have an impressive theory of how these abundances arise, that does not introduce any additional parameters beyond those of the standard models.   One synthesizes a few light isotopes ($H, H^2, He^3, He^4, Li^7$) during the big bang, and the heavier isotopes in stellar burning.   

\subsubsection{Matter, Neutrino, and Densities}

There are many ideas for how an asymmetry between matter and antimatter, which
after much mutual annihilation will boil down to the present baryon
density, might
be generated in the early Universe.  Several of them seem
capable of giving the observed value. Unfortunately the answer
generally depends
on details of particle physics at energies that are unlikely to be
accessible experimentally any time soon. So for a decision among the models we may be
reduced to waiting for a functioning Theory of (Nearly)
Everything.

Similar remarks apply to the neutrino density, as we've already discussed.

I'm much more optimistic about the dark matter problem.  Here we have
the unusual
situation that there are two good ideas.  The symmetry of the standard
model can be enhanced, and some of its aesthetic shortcomings can be overcome,
if we extend it to a larger theory.  Two proposed extensions,
logically independent of one another, are particularly specific and compelling.
One of these incorporates a symmetry suggested by Roberto Peccei
and Helen Quinn \cite{PQ}. PQ symmetry rounds out the logical structure of QCD, by
removing QCD's potential to support strong violation of time-reversal
symmetry, which is not observed.  This extension predicts the existence of
a remarkable new kind of very light, feebly interacting particle:  axions \cite{axions}.
The other incorporates supersymmetry, an extension of special relativity
to include quantum space-timed transformations. Supersymmetry
serves several important qualitative and quantitative purposes in modern
thinking about unification, relieving difficulties with understanding why
W bosons are as light as they are and why the couplings of the standard
model take the values they do. In many implementations of
supersymmetry the lightest supersymmetric particle, or LSP, interacts
rather feebly with ordinary matter (though much more strongly than do
axions) and is stable on cosmological time scales.

The properties of both particles, axion or LSP, are consistent with what we know about dark matter.  Moreover you can calculate how abundantly they would be produced in the
Big Bang.  In both cases the prediction for the abundance is quite
promising.  (I'll discuss the axion calculation further below.)  There are vigorous, heroic experimental searches underway
to dark matter in either of these forms. We will also get crucial
information about supersymmetry, positive or negative, from the Large Hadron Collider (LHC)
starting in 2008. I will be disappointed -- and surprised -- if we don't have
much more specific ideas about the dark matter in hand within a few years.

%%%%%%%%%%%%%%%

\subsubsection{Dark Energy}

The dark energy appeared earlier in another guise, as the fundamental constant $\lambda$ of the (extended) standard model.   From that perspective, it appeared as the intrinsic action per unit volume of space-time.    Several different physical effects should generate contributions to that action density.   Any scalar condensate potentially contributes, including the chiral condensate of QCD, the Higgs condensate of electroweak symmetry breaking, and hypothetical heavier Higgs condensates associated with unified symmetry breaking.   If there are extra small spatial dimensions, there could be contributions from condensates of many more fields, including vectors (fluxes) and gravitons (curvature) with indices in the extra directions.   There could also be an intrinsic, ``bare'' term.  There could also be a term associated with quantum fluctuations, i.e. zero-point energy density.   Theoretical estimates for each of these contributions, with the exception of the chiral condensate, are wildly uncertain.   Notoriously,  ``natural'' expectations, based on  dimensional analysis, for the magnitudes involved are many orders of magnitude larger than the observed net answer.    Either our framework for understanding gravity, even at low energies, is profoundly misleading, or severe cancellations must take place.  

Thus far attempts to augment or modify our theory of gravity in such a way as to make the effective smallness of $\lambda$ appear natural have not led to success.   

\subsubsection{Anthropic Reasoning for Dark Energy}

Several physicists have been led to wonder
whether it might be useful, or even necessary, to take a different
approach \cite{uniOrMulti}.  They invoke a form of observer bias, or anthropic reasoning.  

Here I will first sketch the basic argument in an extreme and oversimplified form.  At the end of this essay, I'll mention some questions and possible objections it raises.

The essential premise of the argument is that many different effective universes exist, with different values of physical parameters (including not only fundamental constants in our sense but also discrete parameters including the gauge groups, their representations among lower-spin fields, the number of families, and the spatial dimension).  By effective universes we might simply mean very distant parts of space, to which we have not so far had observational access; or different components of a universal wave function, with which we have poor overlap; or both.   In any case, it becomes fruitless to try to calculate unique values for the physical parameters, because they are in fact different elsewhere.  We can gain some partial insight into their values -- and test the plausibility of the framework -- by concentrating on the most likely effective universes.   To assess which universes are most likely, we must establish a probability distribution.   In doing that, it is appropriate to use {\it relative\/} probabilities, taking observer bias into account.   That is, what we should try to calculate is not some idealized ``absolute'' or ``God's eye'' expectation value for the parameters, but one that takes into account that the likelihood of observed parameter values is conditioned by the likelihood of sentient observers emerging through physical processes within the relevant domain.   In particular, we must demand that complex structures can emerge.   

Too large a magnitude, positive or negative, of the dark energy will lead to universes that expand or collapse too rapidly for significant structure to form through the normal mechanism of gravitational instability.  Thus only a relatively small value of the net dark energy can be observed.  Indeed, if we hold all other parameters fixed, and let the value of $\lambda$ vary, we seem to find that significantly (say more than 10, or certainly 1000, times) larger values are excluded.    

%%%%%%%%%%%%%%%%%

\subsection{Inflation}

Several assumptions in the standard cosmological model,
specifically uniformity, spatial flatness, and the scale invariant, Gaussian,
adiabatic (Harrison-Zeldovich) spectrum, were originally suggested on
grounds of simplicity, expediency, or esthetics.  They can be supplanted
with a single dynamical hypothesis: that very early in its history the
Universe underwent a period of superluminal expansion, or inflation \cite{linde}.
Such a period could have occurred while a matter field that was coherently
excited out of its ground state permeated the Universe. Possibilities of
this kind are easy to imagine in models of fundamental physics. For example
scalar fields are used to implement symmetry breaking even in the
standard model, and such fields can easily fail to shed energy
quickly enough to
stay close to their ground state as the Universe expands. Inflation
will occur if the approach to the ground state is slow enough.
Fluctuations are
generated because the relaxation process is not quite synchronized
across the Universe.

Inflation is a wonderfully attractive, logically compelling idea, but
its foundations remain amorphous.  Can we be specific about the cause of inflation,
grounding it
in specific, well-founded, and preferably beautiful models of
fundamental physics?  Concretely, can we calculate the correct amplitude of
fluctuations convincingly?  Existing implementations actually have a
problem here; it takes some nice adjustment to get the amplitude
sufficiently small.

More hopeful, perhaps, than the difficult business of extracting hard
quantitative
predictions from such a broadly flexible idea as inflation, is to follow up on the
essentially new and
surprising possibilities it suggests.  The violent restructuring of
space-time attending inflation should generate detectable gravitational waves.
These can be detected through their effect on polarization of the
microwave background. And the non-trivial dynamics of relaxation should
generate some detectable deviation from a strictly scale-invariant
spectrum of fluctuations. These are very well posed questions, begging for
experimental answers.

%%%%%%%%%%%%%%%%%%%

\subsection{Axions and Cosmology}

Given its extensive symmetry and the tight structure of relativistic quantum
field theory, the definition of QCD only requires, and only permits, a very
restricted set of parameters. These consist of the coupling constant and
the quark masses, which we've already discussed, and one more -- the so-called
$\theta$~parameter.  Physical results depend periodically upon
$\theta$, so that effectively it can take values between $\pm \pi$.
We don't know the actual value of the
$\theta$ parameter, but only a limit, $|\theta | \lesssim 10^{-9}$.
Values outside this small range are excluded by experimental results,
principally the tight bound on the electric dipole moment of the neutron. The
discrete symmetries P and T are violated unless $\theta \equiv 0$ (mod $~\pi$).
Since there are P and T violating interactions in the world, the
$\theta$ parameter can't be set to zero by any strict symmetry assumption.
So understanding its smallness is a challenge.

The effective value of $\theta$ will be affected by dynamics, and in particular
by spontaneous symmetry breaking. Peccei and Quinn discovered that if one
imposed a certain asymptotic symmetry, and if that symmetry were
broken spontaneously, then an effective value $\theta \approx 0$
would be obtained.
Weinberg and I explained that the approach
$\theta \rightarrow 0$ could be understood as a relaxation process,
whereby a very
light field, corresponding quite directly to $\theta$, settles into
its minimum energy state.
This is the axion field, and its quanta are called axions.

The phenomenology of axions is essentially controlled by one
parameter, $F$.  $F$ has
dimensions of mass. It is the scale at which Peccei-Quinn symmetry breaks.

\subsubsection{Cosmology}

Now let us consider the cosmological implications.
Peccei-Quinn symmetry is unbroken at temperatures $T\gg F$.
When this symmetry breaks the initial value of the phase is random beyond the then-current horizon scale. One can
analyze the fate of these fluctuations by solving the equations for a
scalar field in an expanding Universe.

The main general results are as follows.  There is an effective cosmic
viscosity, which
keeps the field frozen so long as the Hubble parameter
$H \equiv \dot{R} /R \gg  m$, where $R$ is the expansion factor and $m$ the axion mass.
In the opposite limit
$H \ll m$ the field undergoes lightly damped oscillations, which result in
an energy density that decays as $\rho \propto 1/R^3$.  Which is to say,
a comoving volume contains a fixed mass. The field can be regarded as a
gas of nonrelativistic particles in a coherent state, i.e. a Bose-Einstein condensate. There is some
additional
damping at intermediate stages. Roughly speaking we may say that
the axion field, or any scalar field in a classical regime, behaves
as an effective
cosmological term for $H>>m$ and as cold dark matter for $H\ll m$.
Inhomogeneous perturbations are frozen in while their length-scale exceeds
$1/H$, the scale of the apparent horizon, then get damped as they enter the horizon.

If we ignore the possibility of inflation, then there is a unique
result for the cosmic
axion density, given the microscopic model. The criterion $H \sim m$
is satisfied for
$T\sim \sqrt {\frac{M_{\rm Planck}}{F}}
\Lambda_{\rm QCD}$. At this point the horizon-volume contains many
horizon-volumes
from the Peccei-Quinn scale, but it still contains only a negligible
amount of energy
by contemporary cosmological standards. Thus in comparing to
current observations, it is appropriate to average over the starting
amplitude $a/F$ statistically.
If we don't fix the baryon-to-photon ratio, but
instead demand spatial flatness, as inflation suggests we should,
then for $F > 10^{12}$ GeV the
baryon density we compute is smaller than what we
observe.

If inflation occurs before the Peccei-Quinn transition, this analysis
remains valid.
But if inflation occurs after the transition, things are quite different.

\subsubsection{Inflationary Axion Cosmology: A mini-Multiverse with Controlled Anthropic Reasoning}

For if inflation occurs after the transition, then the
patches where $a$ is approximately homogeneous get
magnified to enormous size. Each one is far larger than the presently
observable Universe.
The observable Universe no longer contains a fair statistical sample of
$a/F$, but some particular ``accidental" value.  Of course there
is still a larger structure, which Martin Rees calls the Multiverse,
over which the value varies.

%%%%%%% for proofs: in next paragraph, fraction was inverted!

Now if $F>10^{12}$ GeV, we could still be consistent with
cosmological constraints on the axion density, so long as
the amplitude satisfies
$ (a/F )^2 \sim ( 10^{12}~{\rm GeV})/F$.  The actual value of $a/F$,
which controls
a crucial regularity of the observable Universe, is contingent in a very
strong sense. Indeed, it is different ``elsewhere".

Within this scenario, the anthropic principle is demonstrably correct
and appropriate.
Regions having large values of $a/F$, in which axions by far dominate
baryons, seem likely to prove inhospitable for the development of complex
structures. Axions themselves are weakly interacting and essentially
dissipationless, and they dilute the baryons, so that these too stay dispersed.
In principle laboratory experiments could discover axions with $F >
10^{12}$ GeV.
If they did, we would have to conclude that the vast bulk of the
Multiverse was inhospitable to intelligent life. And we'd be forced to appeal
to the anthropic principle to understand the anomalously modest axion
density in our Universe.

Even if experiment does not make it compulsory, we are free to analyze the cosmological consequences of $F >> 10^{12}$ GeV .   Recently Tegmark, Aguirre, Rees and I carried out such an analysis \cite{tegmark2}.  We concluded that the although the overwhelming {\it volume\/} of the Multiverse contains a much higher ratio of dark matter, in the form of axions, to what we observe, the typical {\it observer\/} is likely to see a ratio similar to what we observe.

\subsubsection{Dynamical Masses, More Generally?}

Through the anomaly equation, the $\theta$ parameter is connected with the overall {\it phase\/} of the quark mass matrix; indeed, the argument of the determinant of that matrix appears as an additive contribution to $\theta$.   As we discussed earlier, it is through fermion mass matrices that fundamental constants proliferate.  While pattern-seeking eyes can find method in the madness of the observed matrices, certainly the most obvious feature we've recognized is the smallness of that phase (or, to be more precise, its accurate cancellation against other contributions to $\theta$, {\it e.g}. a bare gluon term). The essence of
the Peccei-Quinn mechanism is to promote the phase of quark mass
matrix to an independent, dynamically variable field. Could additional aspects
of the quark and lepton mass matrices likewise be represented as
dynamical fields?  In fact, this sort of set-up appears quite naturally in
supersymmetric models, under the rubric ``flat directions" or ``moduli".
Under certain not entirely implausible conditions particles associated
with these moduli fields could be accessible at future accelerators,
specifically the LHC. If so, their study open a window admitting new light into the
family/Higgs
sector, where we need it badly.

%%%%%%%%%%%%%%%%%%%%%%%%

\section{Questioning Fundamental Constants}

\subsection{Are They Fundamental?}

Over the course of this essay, we've seen many reasons to question whether the fundamental constants that appear in our present standard model of physics are truly fundamental in the ordinary sense of the word.  

\subsubsection{Reductions?}

On the one hand, several calculations and ideas suggest ways to analyze some of the quantities we presently regard as fundamental constants into more basic elements: 
\begin{itemize}
\item The gauge couplings appear ripe for unification.  From that perspective, two parameters -- the unification mass scale, and the coupling at unification -- appear more fundamental than the three independent couplings of the standard model.   An objective sign of this is that there are fewer of them!
\item The unification plausibly, and semi-quantitatively, extends to gravity.   
\item Neutrino masses and mixings plausibly arise indirectly, through processes associated with unification.  
\item The small value of the $\theta$ parameter is plausibly explained by Peccei-Quinn symmetry, leading to the physics of axions.
\end{itemize}
This reductive theme extends to the parameters of the standard model of cosmology:
\begin{itemize}
\item The density of neutrinos and of baryons plausibly reflects microscopic physics at large energy scales, playing out through the big bang.
\item Both the character of dark matter -- i.e. the fundamental constants governing {\it its\/} behavior -- and its density should be traced to microscopic physics.   If either the lightest supersymmetric partner or the axion contributes significantly to the dark matter, we have realistic prospects for major progress on this question in the near future.  
\item Inflation might be traced to phase transitions or field evolution, again in the context of the big bang.
\item The amplitude of fluctuations might be calculable within a specific, microscopically grounded realization of inflation. 
\end{itemize}

\subsubsection{Selections and Accidents?}

On the other hand, other fundamental constants have so far resisted theoretical elucidation:
\begin{itemize}
\item The value of $\lambda$, the dark energy parameter, appears as the sum of diverse contributions, each with large magnitude, that very accurately cancel.
\item Similar remarks apply to the value of $\mu^2$, the Higgs mass parameter, or to the semi-equivalent mass parameters that appear in non-minimal extensions of the standard model.   In models with low-energy supersymmetry some of the required cancellations occur automatically.  
\item The many Higgs coupling parameters, which reflect themselves in quark and lepton masses and mixing angles, have utterly resisted calculation.   At best, the number of parameters might be reduced by approximate (perhaps spontaneously broken) flavor symmetries.
\end{itemize}

The early promise of superstring theory to calculate these quantities has faded after decades of disappointing experience in attempts to construct phenomenologically adequate solutions, together with the discovery of multitudes of theoretically unobjectionable but empirically incorrect solutions.  

At the same time, the success of inflationary cosmology has made it increasingly plausible that the universe we presently observe, wherein the same fundamental parameters seem to hold everywhere, might be only part of a much larger structure.   Within that larger multiverse, the fundamental parameters could vary.   

Together, these developments suggest the possibility that many many solutions of the basic equations are in fact realized in different parts of the multiverse.   If so, attempts to calculate the values of all the fundamental constants from the microscopic theory are doomed, since those ``constants'' actually take on different values in different places.    

Nothing in this scenario precludes that {\it some\/} of the values of fundamental constants might reflect profound symmetries and dynamics of physics beyond the standard model.   Indeed, in the preceding paragraphs we've identified several promising cases.   But others might not be.

Among the fundamental constants that are not constrained by profound principles, a few might be constrained by selection effects, as we've discussed in the case of dark energy.    Others might have values that are essentially accidental.   It is difficult to imagine, for instance, that the precise mass of the bottom and charm quarks, or their mixing angles, has much impact on the evolution of sentient observers.   

There is nothing logically inconsistent in this view of the world.  Indeed, the ``unnatural'' fine tunings of some fundamental parameters required for the existence of life has long been remarked, as has the irrelevance of others, and our general failure in either case to find profound symmetry or dynamical meaning in their specific values.   

At the moment, however, we do not know which combinations of fundamental constants vary over the hypothetical multiverse, and which are constrained by powerful principles.   As an example of the importance of that question, consider the case of dark energy.  If the dark energy density is assigned a flat {\it a priori\/} probability distribution, with all other fundamental constants and cosmological parameters held fixed, then the value we observe is not unlikely, based on selection through observer bias.   But if we allow both the dark energy density and the amplitude of cosmic fluctuations to vary, the story is very different.   Universes with much more dark energy than we observe, and larger fluctuation amplitudes, become quite likely.    

Absent  profound microscopic understanding, anthropic arguments will always be subject to objections of this kind.   Inflationary axion cosmology appears to be a uniquely favorable case, where the microphysical hypotheses are clearly formulated and cleanly related to a cosmological parameter (that is, the dark matter density).

\subsection{Are They Constant?}

Finally: If the values of fundamental constants vary from place to place, they might also be expected to evolve in time.   If different effective universes differ discretely, and are separated by large energy barriers, transitions might be very rare and catastrophic.   But if there are light fields that vary continuously, their evolution might manifest itself as an apparent change in the fundamental constants.  Thus for example changes in the value of a scalar field $\eta$ that couples to the photon in the form
${\cal L}\ \propto \eta F_{\mu \nu} F^{\mu \nu}$ would appear as changes in the value of the fine structure constant.

\end{document}